# Fast and direct nonparametric procedures in the L-moment homogeneity test

Pierre Masselot[1*], Fateh Chebana[1], Taha B.M.J. Ouarda[1,2]

February, 2015

[1]Centre Eau-Terre-Environnement (ETE), Institut national de la recherche scientifique (INRS), 490 de la Couronne, Québec, QC, G1K 9A9, Canada

[2] Institute center for water and environment (iWATER), Masdar institute of science and technology, PO Box 54224, Abu Dhabi, United Arab Emirates

[*] Corresponding author : pierre-lucas.masselot@ete.inrs.ca

## Abstract

Regional frequency analysis is an important tool to properly estimate hydrological characteristics at ungauged or partially gauged sites in order to prevent hydrological disasters. The delineation of homogeneous groups of sites is an important first step in order to transfer information and obtain accurate quantile estimates at the target site. The Hosking-Wallis homogeneity test is usually used to test the homogeneity of the selected sites. Despite its usefulness and good power, it presents some drawbacks including the subjective choice of a parametric distribution for the data and a poorly justified rejection threshold. The present paper addresses these drawbacks by integrating nonparametric procedures in the L-moment homogeneity test. To assess the rejection threshold, three resampling methods (permutation, bootstrap and Pólya resampling) are considered. Results indicate that permutation and bootstrap methods perform better than the parametric Hosking-Wallis test in terms of power as well as in time and procedure simplicity. A real-world case study shows that the nonparametric tests agree with the HW test concerning the homogeneity of the volume and the bivariate case while they disagree for the peak case, but that the assumptions of the HW test are not well respected.

## Acknowledgements

The data used were simulated using Matlab computer codes available upon request from the authors.

# 1. Introduction

Frequency analysis (FA) of extreme events, such as floods and droughts, has major implications in the field of water resources management. FA aims to estimate the risk associated to these events through the fitting of a probability distribution to observed data (*e.g.* Chow et al. 1988; Lekina et al. 2013). However, regional FA (RFA) is required to estimate similar risk at sites where little or no hydrologic data is available. In RFA, regions are formed by grouping sites with similar hydrologic features and then hydrological information can be transferred to those target sites (Hosking and Wallis 1997). RFA is now widely used and is still an active research topic (*e.g.* Bates et al. 1998; Burn and Goel 2000; Lin and Chen 2006; Ribatet et al. 2006; Seidou et al. 2006; Haddad and Rahman 2012; Zaman et al. 2012; Ouali et al. 2015; Wazneh et al. 2015; Wright et al. 2015). The key assumption of RFA is the homogeneity of the sites forming a region regarding their hydrological distribution. If the region is not homogeneous, it could affect the accuracy of the estimation at ungauged sites (e.g. Lettenmaier et al. 1987; Chebana and Ouarda 2008). Thus, specific attention must be given to ensure regional homogeneity.

Over the years, several statistical tests have been developed to decide if a set of sites forms a homogeneous region (*e.g.* Wiltshire 1986; Lu and Stedinger 1992; Hosking and Wallis 1993; Fill and Stedinger 1995). Among those tests, the homogeneity test of Hosking and Wallis (1993) (denoted HW test in the following) has established as the reference in the literature. It is indeed usually used and continues to be used in practice (*e.g.* Fowler and Kilsby 2003; Norbiato et al. 2007; Cannarozzo et al. 2009; Ngongondo et al. 2011; Núñez et al. 2011; Santos et al. 2011; Lim et al. 2013) and to be considered in statistical developments (*e.g.* Castellarin et al. 2008; Das and Cunnane 2010; Chérif and Bargaoui 2013; Wazneh et al. 2015; Wright et al. 2015). The HW test has recently been extended to the multivariate case by Chebana and Ouarda (2007) to simultaneously take into account the various characteristics of flood events (peak, volume and duration). Note that the hypothesis of a regional distribution represents the foundation of the Index-Flood method (see for instance Dalrymple (1960), Hosking and Wallis (1997) or Chebana and Ouarda (2009)).

Given some attractive features of the HW test and its popularity in RFA, the present paper focuses on this test in order to avoid some of its drawbacks and also to be considered as a benchmark. This test is based on the L-moments (Hosking 1990) which are robust to extreme values and allow to uniquely define a distribution (see Hosking and Wallis (1997), chapter 2, for a discussion of their properties). Even though the HW test has a good power (e.g. Viglione et al. 2007), it still suffers from two main drawbacks. Firstly, its application requires the fitting of a parametric distribution to the data (although flexible) leading to the estimation of its parameters and secondly, the rejection threshold is not well justified. The subjective choice by Hosking and Wallis (1993) of the four-parameters Kappa distribution and the estimation of its parameters create (unnecessary) uncertainty in the estimated test statistic distribution and then in the estimated rejection threshold. From a conceptual point of view, it seems inefficient that a test requires modeling aspects which are posterior to this test. Indeed, a parametric distribution is fitted to data in order to test for a homogeneous region and once this region is formed, a parametric distribution is fitted to its data. Furthermore, parameter estimation is achieved through numerical optimization methods which do not always converge. Finally, because of the lack of theoretical justification of the rejection threshold, the decision is not straightforward when the value of the test statistic is close to the threshold. These drawbacks can cause unstable results and complicate the application for the practitioners. Indeed, the homogeneity test is a preliminary step in the RFA process and should be as simple as possible (Hosking and Wallis 1997). Note that these drawbacks become more important in the multivariate case where the number of parameters increases rapidly with the number of studied variables.

The present paper proposes to overcome the drawbacks of the HW test by providing a nonparametric framework in the calculation of the rejection threshold. The purpose is to avoid the unnecessary step of fitting a parametric distribution to the data. This increases the accuracy and the applicability of the test. The proposed approach consists in replacing the parametric Monte Carlo simulations by nonparametric methods, namely the permutation methods (Fisher 1935; Pitman 1937), the bootstrap (Efron 1979) or the Pólya resampling (Lo 1988). Although it is well known that a parametric test is often more powerful, it requires assumptions about the underlying distribution of data which are not always met in practice. A nonparametric framework avoids dealing with too many assumptions, and increases the applicability and the performance of the test when the assumptions are not met (Good 2005, p.47). For instance, Viglione et al. (2007) concluded that, for highly skewed regions, the HW test is outperformed by the bootstrap based k-sample Anderson-Darling test of Scholz and Stephens (1987). According to Hirsch et al. (1991), nonparametric methods offer modest disadvantages when data fits well a known distribution but large advantages when the underlying data distribution is unknown. This is especially true for RFA considering the homogeneity test represents a preliminary step prior to the fitting of a distribution to hydrologic data. Furthermore, it is proposed to take the decision of rejecting the homogeneity according to a p-value instead of fixed threshold in the HW heterogeneity measure. This provides a theoretically justified and automatic decision to the test. Thus, the new test does not need user intervention except fixing a significance level (usually 5%). Note that the proposed improvements to the test hold equally for both univariate and multivariate settings since the latter generalizes the former.

The organisation of the paper is as follows. Section 2 briefly presents the necessary background related to the L-moments and the original HW test. The proposed nonparametric test and the nonparametric methods used are detailed in section 3. The HW test and the proposed alternatives are compared in a simulation study in section 4 and for a real world application in section 5. The last section concludes the study.

## 2. Background

This section briefly presents the tools needed to develop the homogeneity test. The L-moments used to define sites' distribution are first introduced followed by the HW test. The test is introduced with its multivariate version (Chebana and Ouarda 2007).

### 2.1. L-moments

L-moments were introduced by Hosking (1990) as an alternative to traditional moments and have been extended to the multivariate case by Serfling and Xiao (2007). They are often used in hydrology because of their interesting properties for modeling heavy-tailed distributions. Thus, the HW test statistic is naturally based on L-moments. This section briefly presents univariate and multivariate L-moments to describe the shape of probability distributions. Univariate L-moments are presented in details in Hosking and Wallis (1997) with properties and advantages.

L-moments are a linear combination of order statistics sampled from an underlying distribution. The univariate L-moment of order $r$ is defined as

$$\lambda_r = \int_0^1 F^{-1}(u) P_{r-1}^*(u) du \qquad (1)$$

where $P_{r-1}^*(u)$ is the so-called shifted Legendre polynomial (e.g. Chang and Wang 1983) and $F(\cdot)$ is the cumulative distribution function (cdf) of the variable of interest.

The $r^{th}$ multivariate L-moment is a $p \times p$ matrix $\Lambda_r$ containing at line $i$ and column $j$, by analogy with a covariance matrix, the L-comoment defined by :

$$\lambda_{r[ij]} = \text{cov}\left(X^{(i)}, P_{r-1}^*\left(F_j\left(X^{(j)}\right)\right)\right) \quad (2)$$

Note that, unlike a covariance matrix, the elements $\lambda_{r[ij]}$ and $\lambda_{r[ji]}$ are not necessarily equal. Multivariate L-moments capture the behavior and the attractive properties of univariate L-moments (Serfling and Xiao 2007).

To summarize a distribution, it is convenient to define dimensionless versions of L-moments (Hosking and Wallis 1997). It is achieved by defining the L-moment ratios

$$\tau_r = \frac{\lambda_r}{\lambda_2}, \text{ for } k \geq 3 \text{ and } \tau = \frac{\lambda_2}{\lambda_1} \quad (3)$$

The latter quantity $\tau$ is called L-CV, and the quantities $\tau_3$ and $\tau_4$ are respectively called L-skewness and L-kurtosis with obvious analogues. The multivariate cases are simply matrices $\Lambda_r^*$ with L-comoment coefficients given by

$$\tau_{r[ij]} = \frac{\lambda_{r[ij]}}{\lambda_2^{(i)}}, \text{ for } k \geq 3 \text{ and } \tau_{2[ij]} = \frac{\lambda_{2[ij]}}{\lambda_1^{(i)}} \quad (4)$$

where $\lambda_r^{(i)}$ is the univariate $r^{th}$ L-moment of the variable $X^{(i)}$.

The estimators of the L-moments on finite samples are presented by Serfling and Xiao (2007).

## 2.2. The Hosking-Wallis (HW) test

In the following, the letters "HW" refer to the multivariate version of the test (Chebana and Ouarda 2007) since it is more general and includes the univariate version as a special case.

As explained in the introduction, the HW test aims to reject or not the hypothesis of homogeneity of a set of $N$ sites, according to appropriate hydrological variables. Hosking and Wallis (1997) indicated that, in a homogeneous region, the sites frequency distributions have to be the same except for a site-specific scale factor. This is why the test statistic is based on dimensionless characteristics such as the L-ratios $\Lambda_r^*$ (or $\tau_r$ for univariate data). More precisely, the test statistic is based on the L-CV, $\Lambda_2^*$ only since it is the most critical characteristic to determine the accuracy of RFA (Hosking and Wallis 1997, section 7.5.7.). Note also that Viglione et al. (2007) concluded that the inclusion of the L-skewness $\tau_3$ decreases the power of the HW test. The test statistic is thus defined as

$$V = \left( \frac{\sum_{j=1}^{N} n_j \left\| \Lambda_2^{*(j)} - \overline{\Lambda_2^*} \right\|^2}{\sum_{j=1}^{N} n_j} \right)^{\frac{1}{2}} \tag{5}$$

where $n_j$ $(j=1,...,N)$ is the sample size of the $j^{\text{th}}$ site, and $\overline{\Lambda_2^*} = \left( \sum_{j=1}^{N} n_j \right)^{-1} \sum_{j=1}^{N} n_j \Lambda_2^{*(j)}$ is the weighted mean of at-sites' L-CVs $\Lambda_2^{*(j)}$. Any matrix norm can be applied to compute $V$ but Chebana and Ouarda (2007) retained the "spectral" norm (defined as $\|A\|_2 = \sqrt{\text{maximum eigenvalue of } A^t A}$) as the most appropriate in this case. Note that $V$ is the weighted standard deviation of the at-site L-CVs. This induces that, intuitively, this test can be seen as a "kind of ANOVA" since the variance between groups is tested. Note also that, for the case of a single variable, matrices $\Lambda_2^{*(j)}$ reduce to scalar $\tau_2^j$ values, which results in the $V$ statistic of Hosking and Wallis (1993).

To decide if the region can be considered homogeneous, $V$ has to be compared with some reference values representing the expected values of $V$ for an homogeneous region. This is achieved by simulating $N_{sim}$ homogenous regions $\mathbf{X}^{(b)}$ $(b=1,...,N_{sim})$ of the same at-site size as the observed region in order to compute $N_{sim}$ replicates $V^{(b)}$. For each site, margins are simulated from a four-parameter Kappa distribution and the dependence between variables is simulated from an extreme value copula using the algorithm of Ghoudi et al. (1998). The parameters needed for the simulation of each site are fitted on the pooled distribution of all sites. Once the $V^{(b)}$ values are computed, the "heterogeneity measure" is defined as

$$H = \frac{V - \mu_{sim}}{\sigma_{sim}} \qquad (6)$$

where $\mu_{sim}$ and $\sigma_{sim}$ are respectively the mean and the standard deviation of the $V^{(b)}$ values. As a rule of thumb, the region is considered homogenous if $H < 1$, possibly homogeneous if $1 < H < 2$ and heterogeneous otherwise. Assuming $V$ is normally distributed, one can also reject the hypothesis of homogeneity at a significance level of 5% when $H > 1.64$ (Fill and Stedinger 1995). The whole HW procedure is summarized in Algorithm A.

---

**Algorithm A**: Algorithm of the Hosking-Wallis homogeneity test

Let $\mathbf{X}^{obs} = \{\mathbf{x}_1^{obs},...,\mathbf{x}_N^{obs}\}$ be a set of observed sites (a region) of respective sizes $n_1,...,n_N$.

A1. Compute $V^{obs}$ using (5) on the observed sites ;

A2. Simulate a large number of homogeneous regions through generating $\mathbf{X}^{(b)} = \{\mathbf{x}_1^{(b)},...,\mathbf{x}_N^{(b)}\}$ $b=1,...,N_{sim}$ with the sites record length $(n_1,...,n_N)$ using the Kappa distribution and the extreme value copula fitted on $\mathbf{x}^{tot} = \bigcup_{j=1}^{N} \mathbf{x}_i^{obs}$ ;

A3. Compute $V^{(b)}$ using (5) on each of the simulated regions $\mathbf{X}^{(b)}$ ;

A4. Compute the heterogeneity measure $H$ using (6) and decide that the region is homogenous if $H<1$, acceptably homogeneous if $1<H<2$ and heterogeneous when $H>2$, or reject the hypothesis of homogeneity at the 5% significance level when $H>1.64$.

## 3. The proposed nonparametric test

This section presents the chosen solutions to overcome the drawbacks of the HW test. In the following, the test statistics obtained within the nonparametric framework are denoted using a "*" to mark the difference with the original parametric framework. The improvements are made on the most important steps A2 and A4 of the algorithm A.

First, the main idea of the new test is to substitute the parametric Monte Carlo simulation of step A2 by a nonparametric one. This means the new regions $\mathbf{X}^{*(b)}$ $(b=1,...,N_{sim})$ are generated on the basis of the empirical distribution of $\mathbf{X}^{obs}$. This is intuitively simpler and more automatic for the user. To obtain $\mathbf{X}^{*(b)}$ by nonparametric simulations, three methods are presented here : the permutation methods (Fisher 1935; Pitman 1937), the bootstrap (Efron 1979) and the Pólya resampling (Lo 1988). The three methods are explained below in this section and compared in the simulation study of section 4.

Second, once the $\mathbf{X}^{*(b)}$ are simulated and the $V^{*(b)}$ computed, the heterogeneity measure of step A4 is replaced by a rejection threshold based only on the $V^{*(b)}$s. Since the test is one-sided, the threshold for a significance level $\alpha$ is $v_{1-\alpha}$ which is the quantile of order $1-\alpha$ of the set of $V^{*(b)}$ values. The corresponding p-value is given by (defined in Efron and Tibshirani (1993, chapter 16)):

$$p-value_H = N_{sim}^{-1} \#\left(V^{*(b)} > V^{obs}\right) \qquad (7)$$

where the homogeneity is rejected when $p-value_H < \alpha$. The nonparametric procedure is summarized in Algorithm B with new steps B2 and B4.

---

**Algorithm B**: Algorithm of the Nonparametric homogeneity test

Let $\mathbf{X}^{obs} = \left\{\mathbf{x}_1^{obs},...,\mathbf{x}_N^{obs}\right\}$ be a set of observed sites (a region) of respective sizes $n_1,...,n_N$. Choose a significance level $\alpha \in [0;1]$.

B1. Compute $V^{obs}$ using (5) on the observed sites ;

B2. Simulate a large number of homogeneous regions through generating $\mathbf{X}^{*(b)} = \{\mathbf{x}_1^{*(b)},...,\mathbf{x}_N^{*(b)}\}$ $b=1,...,N_{sim}$ with at-sites record length ($n_1,...,n_N$) as $\mathbf{X}^{obs}$ by sampling from the *empirical* distribution of $\mathbf{x}^{tot} = \bigcup_{j=1}^{N} \mathbf{x}_i^{obs}$;

B3. Compute $V^{*(b)}$ using (5) on each of the simulated regions $\mathbf{X}^{*(b)}$;

B4. Compute the $p-value_H$ using (7). The region is considered heterogeneous with an error $\alpha$ if $p-value_H < \alpha$.

The following subsections present the three possibilities for step B2 which are also illustrated in Figure 1.

### 3.1. The permutation (M) test

Permutations refer to the construction of new data sets by permuting observed data, *i.e.* by changing the labelling of each observation. In the RFA context, the corresponding test (called M test) simulates homogeneous regions $\mathbf{X}^{*(b)}$ by reassigning randomly each observation to a given site. The idea behind such an approach is that, if the null hypothesis of homogeneity is true, then values from a given site could occur in any other site. The observed region is thus one of the possible permutations of observed data between sites.

Theoretically, the test statistic $V$ should be computed for each of the possible data permutations to estimate its underlying distribution under the null hypothesis. However, since the number of possible permutations increases extremely fast with the sample size (for example, there are more than eleven billion possibilities of rearranging twenty data in four samples of size 5), it is impossible to obtain the whole permutation distribution, even for modern computers. This is why, in practice, only a subset of all possible permutations is considered (Ernst 2004). In other words, each simulated site $\mathbf{x}_j^{*(b)}$ is drawn by sampling *without* replacement from the pooled sample of all sites $\mathbf{x}^{tot}$.

Permutation methods can be traced back to Fisher (1935) and Pitman (1937)where it was developed to apply the t-test when data are not Gaussian. Permutation methods are mainly applied to test homogeneity or equality hypotheses (e.g. Ernst 2004) although more complex ones can be tested this way (e.g. Welch 1990). The main advantage of permutation methods is that the probability of making a type I error is theoretically equal to the chosen significance level $\alpha$ (Ernst 2004). A test having this property is said to be "exact". This is a desirable property since it allows a precise control over the probability of being wrong in rejecting the null hypothesis Good (2005, p.22).

### 3.2. The bootstrap (B) test

The bootstrap is a computer intensive method of statistical inference introduced by Efron (1979). Like permutation methods, the Bootstrap is fully data driven since the inference is made only from available data. The fundamental idea is that the empirical distribution of a variable is an estimation of its true distribution. Since this true distribution is unknown, the empirical distribution becomes the basis to simulate new datasets having the same characteristics. This is done by drawing samples *with* replacement from the reference sample. The B procedure in the RFA context is then: the sites $\mathbf{x}_j^{*(b)}$ are simulated by sampling *with* replacement from the pooled sample $\mathbf{x}^{tot}$. Note that, as explained by Efron and Tibshirani (1993), in the hypothesis testing framework, the bootstrap is closely related to permutation methods, where the former draws a sample *with* replacement while the latter draws *without* replacement. Although not necessarily exact, Bootstrap tests are more widely applicable than permutation tests.

As a bootstrap procedure, some alternatives to the B test can be considered. One can resample each site $\mathbf{x}_j^{*(b)}$ from its own observation set $\mathbf{x}_j^{obs}$ (a "within site bootstrap") or one can also pick at random an integer $k$ corresponding to a different site from $\{1;...;N\}$ *with* replacement and set $\mathbf{x}_j^{*b} = \mathbf{x}_k^{obs}$ (a "between site bootstrap"). However, these two latter procedures do not respect the first recommendation of Hall and Wilson (1991) who indicated that, to be relevant in the hypothesis testing framework, the bootstrap must be done in a way that reflects the null hypothesis even if this null hypothesis is false. In the RFA context, this means that all sites $\mathbf{x}_j^{*(b)}$ of a new region $\mathbf{X}^{*(b)}$ must be drawn from the exact same distribution which could only be the distribution of the pooled sample $\mathbf{x}^{tot}$.

To test the equality of different samples' scale parameters, Boos and Brownie (1989) explained that before being pooled, the data from each sample must be centered. Since RFA is a very similar case (recall that $V$ is constructed in such a way that the at-site L-CVs are compared), applying such an approach is equivalent to drawing $\mathbf{x}_j^{*(b)}$ by sampling from $\tilde{\mathbf{x}}^{tot}$, *i.e.* the pooled sample containing the values

$$\tilde{x}_{ij}^{tot} = x_{ij}^{tot} - m_j + m \qquad (8)$$

where $x_{ij}^{tot}$ is the $i^{th}$ value of the $j^{th}$ observed site with sample mean $m_j$ and $m$ is the overall mean. This allows the test to respect the kurtosis of the observed sites as well as having the exactness property even when the $m_j$ values are actually different (Boos and Brownie 1989). The first condition of Hall and Wilson (1991) of the null hypothesis respect for bootstrap tests is then fully achieved by the centering of pooled values. Nevertheless, the two bootstrap procedures corresponding to sampling from non-centered or centered pooled values are compared in the simulation study of section 4. The two procedures are respectively denoted "B" test and "Bc" test.

### 3.3. The Pólya (Y) test

The bootstrap introduced in the previous section is based on the assumption that the observed sample correctly represents the underlying distribution of data. This is a strong hypothesis, especially when the observed sample has a small record length (Rubin 1981). In this context, a Bayesian counterpart to the bootstrap, useful to assess the missing information, called "Pólya resampling" and based on a Pólya urn scheme, was developed by Lo (1988).

The Pólya resampling is very similar to the bootstrap with the difference that, each time an observation is drawn from the observed sample, this observation is replaced twice in the sample. Therefore, the probability of sampling from an interval increases each time a value from this interval is chosen. Thus, a Pólya sample interpolates the underlying distribution by modifying the probability of observed values. This technique takes into account the sampling variability of the observed data because different Pólya samples would interpolate at different places. The Pólya resampling is in the Bayesian framework since it is equivalent to considering that the probability of drawing $q$ times each value follows a Dirichlet distribution for which parameters change at each step (Lo 1988).

For the hypothesis testing framework, two possibilities are hereby considered. First, each simulated site $\mathbf{x}_j^{*(b)}$ is drawn from $\mathbf{x}^{tot}$ according to the Pólya resampling scheme. The drawback of this first technique is that each sampling interpolates at different locations resulting in different distributions. Therefore, this scheme does not exactly respect the null hypothesis of homogeneity, violating the first condition of Hall and Wilson (1991). Second, a new "observed" sample $\mathbf{x}^{Polya}$ can be drawn from $\mathbf{x}^{tot}$ with the same total record length (*i.e.* $n = \sum_{j=1}^{N} n_j$) using the Pólya resampling. The sites are then drawn from $\mathbf{x}^{Polya}$ using the traditional bootstrap. The advantage is that all the sites are generated form the same distribution. However, the drawback is that it creates another sampling level, resulting in another level of variability. This new level of variability is unfortunately added to the variability of the final threshold (see Davison and Hinkley (1997) for a discussion about sampling levels). These two possibilities, denoted respectively "Ys" and "Yr" are compared in the simulation study of section 4.

## 4. Simulation study

The purpose of the simulation study is to evaluate and compare the performances of the proposed nonparametric procedures (M, B, Bc, Ys and Yr tests, all generally given by the algorithm B) to the original HW test as well as between themselves. Comparisons are made in both univariate and multivariate frameworks. To this end, different hydrological regions are generated in order to compare their true value (homogeneous or not) to the decision taken by the tests.

The design of the simulation study is first explained before presenting the obtained results.

### 4.1. Design

The design of the simulation study is inspired by Hosking and Wallis (1997, chapter 7) concerning the univariate framework and by Chebana and Ouarda (2007) for the multivariate framework. The study follows three steps:

S1. Creation of $P$ artificial regions with identical known characteristics;

S2. Application of the tests on each of the $P$ regions. This results in $P$ decisions for each test;

S3. Evaluation of the performances of each test by computing a rejection rate.

The three steps of the simulation study are detailed in the following.

### 4.1.1. Creation of regions

A region is generated with $N$ sites ($N = 15, 20, 30$) and a fixed record length $n_j = 30$ for each site $j$.

**Univariate regions**

Univariate regions (UR) are generated, as in Hosking and Wallis (1997, chapter 7), according to a lognormal distribution with L-CV $\tau = 0.08$ and L-skewness $\tau_3 = 0.05$. It is a plausible model for positive and slightly skewed hydrological data. To be exhaustive and be able to evaluate the performance according to different practical cases, several kinds of regions are generated for univariate regions:

a. *Homogeneous*: both parameters $\tau$ and $\tau_3$ are the same for all sites of the region.

b. *Heterogeneous linear*: L-CV $\tau$, L-skewness $\tau_3$, or both vary linearly from site 1 to site $N$. The heterogeneity depends on a rate $\gamma \in [0;1]$ meaning that the corresponding parameter varies linearly from $\tau(1-\gamma/2)$ to $\tau(1+\gamma/2)$. Heterogeneity rates for $\tau$ and $\tau_3$ are respectively denoted $\gamma_\tau$ and $\gamma_{\tau_3}$ in the following.

c. *Heterogeneous bimodal*: two groups of sites are defined according to the L-CV $\tau$, L-skewness $\tau_3$, or both. The concerned parameters are different between the two groups but homogeneous within each group. As for linearly heterogeneous regions, the difference between the two groups depends on the heterogeneity rates $\gamma_\tau$ and $\gamma_{\tau_3}$ to set the concerned parameters to $\tau(1-\gamma/2)$ and $\tau(1+\gamma/2)$.

For both linear and bimodal cases, the heterogeneity rates are set to $\gamma_\tau = 37.5\%$ and $\gamma_{\tau_3} = 200\%$ which represent very common heterogeneity levels (Hosking and Wallis 1997, chapter 7). This makes the L-CV $\tau$ vary from 0.065 to 0.095 and the L-Skewness $\tau_3$ vary from 0 to 0.1.

**Multivariate regions**

Multivariate regions (MR) follow the scheme of Chebana and Ouarda (2007) by simulating plausible bivariate models containing the peak and volume of floods. Among the main three variables describing a flood hydrograph (volume, peak and duration), it has been shown that peak and volume are the most correlated (*e.g.* Yue et al. 2001; Grimaldi and Serinaldi 2006; Vittal et al. 2015) and are thus often used in a bivariate analysis (*e.g.* Chebana and Ouarda 2011; Volpi and Fiori 2012; Santhosh and Srinivas 2013). Note that a large volume of at-site frequency analysis literature deals with the bivariate case while few studies deal with the trivariate case. The reason could be the complexity and low representativeness of higher dimension copulas. This issue becomes more problematic in a regional framework and even more in a simulation study. Chebana and Ouarda (2009; 2011) provided a summary of the difficulties that rise in high dimensions.

The margins follow a Gumbel distribution since this distribution has been shown to provide a good representation of flood peaks and volumes (*e.g.* Yue et al. 1999; Shiau 2003). The dependence is modelled using the Gumbel logistic copula. Zhang and Singh (2006) have shown that this copula is superior for modelling the dependence between the peak and volume. However, other distributions and copulas could also be considered, depending on the application. The corresponding parameters of the considered distribution are those obtained from the data series of the Skootamatta basin in Ontario, Canada, *i.e.* the peak variable is considered to follow a Gumbel distribution with a location parameter $\mu_Q = 52$ and a scale parameter $\sigma_Q = 16$ while the volume variable is defined with a location parameter $\mu_V = 1240$ and a scale parameter $\sigma_V = 300$. Moreover, the dependence parameter in the Gumbel copula is set to $m = 1.41$ corresponding to a correlation coefficient between the two variables of $\rho = 0.5$. Here, the different kinds of generated regions are:

a. *Homogeneous*: all parameters are the same for all sites of the region;

b. *Heterogeneous on the marginal parameters*: it is the same as the heterogeneous linear of UR but on the scale parameters $\sigma_Q$ and $\sigma_V$ of the margins;

c. *Heterogeneous on the dependence parameter*: same as above but on the dependence parameter;

d. *Completely heterogeneous*: heterogeneous on both the marginal and dependence parameters;

e. *Bimodal on the marginal parameter*: the same as heterogeneous bimodal of UR but on the scale parameters $\sigma_Q$ and $\sigma_V$ of the margins;

f. *Bimodal on the dependence parameter*: same as above but on the dependence parameter;

g. *Completely bimodal*: bimodal on both the marginal and dependence parameters.

### 4.1.2. Application of the tests

After choosing the type of region for which the tests are evaluated (step S1), a large number of regions of this type are generated. This number is $P=500$ for UR and $P=100$ for MR, because the latter is much more computationally intensive to create and to apply to all the tests. Here, it is important to distinguish $P$ from $N_{sim}$. Indeed, $N_{sim}$ is the number of replications to assess the distribution of $V$ *within* the tests, while $P$ is *outside* the tests and is the number of the generated regions in order to measure the tests performances. Each of the HW, M, B, Bc, Ys and Yr tests are then applied to each of the $P$ generated regions. Note that, in order to compare the results, a p-value, as in step 4 of algorithm B, is computed for the HW test instead of the $H$ measure.

### 4.1.3. Performance evaluation

Once the tests have been applied (step S2) on each of the $P$ generated regions, a rejection rate is computed for each test. Note that this third step of the simulation study is applied regardless of the framework UR or MR. The computed rejection rate can estimate two features of a test. First, if the kind of region being chosen in step S1 is homogeneous, the rejection rate estimates the first type error $\hat{\alpha}$ which is the probability to reject the null hypothesis of homogeneity while the region is actually homogeneous. Theoretically, a powerful test (according to Neyman and Pearson 1933) must have a first type error close to the chosen significance level $\alpha$. Second, if the kind of region chosen in step S1 is heterogeneous (linear or bimodal), the rejection rate estimates the power $\hat{\beta}$ of the test. The power of a test is the probability to make the right decision by rejecting the null hypothesis of homogeneity. The closer to 100% is $\hat{\beta}$, the more powerful the test is, since it means that the test is able to detect departures from the null hypothesis.

Beyond $\hat{\alpha}$ and $\hat{\beta}$, tests can be compared based on their applicability, which means computing time, number of steps and the need for user intervention (subjectivity).

### 4.2. Results

The first desirable property is the exactness. Recall that a test is sais "exact" if its first type error is equal (or close in practice) to the significance level $\alpha$ (Good 2005, p.22). Estimates of the first type error $\hat{\alpha}$ for UR are shown in Table 1. Chebana and Ouarda (2007) reported slightly too high first type errors for the HW test. This result is confirmed by Table 1 where the first type errors are between 6% and 7% for an $\alpha = 5\%$ significance level test. This means that the critical region is too large and that the estimated power could be overestimated. However, M and B tests show first type errors that are close to $\alpha = 5\%$. This is expected for the M test because permutation methods are known to be exact. For the Bc test, *i.e.* when data are centered before resampling, the first type error is always above $\alpha = 5\%$, which means that this test is not exact. On the contrary, the Ys and Yr tests are slightly conservative since they show first type errors between 3.4% and 4.2% which is always beneath $\alpha = 5\%$.

The first type error estimates $\hat{\alpha}$ for MR are shown in Table 2. One can observe that $\hat{\alpha}$ for the multivariate HW test is between 1.0% and 3.2% which is much lower than the significance level $\alpha = 5\%$. This is different of the results shown in the univariate framework and to the results of Chebana and Ouarda (2007) which reported first type errors above 5%. The B, Ys and Yr tests are also conservative. Like the UR framework and as expected, the multivariate M test has first type estimates $\hat{\alpha}$ close to the significance level $\alpha = 5\%$. This is also the case for the Bc test which confirms that it rejects the null hypothesis more easily than the B test (*i.e.* bootstrap when sites are not centered). This shows that the centering of data before resampling helps simulating actual homogeneous regions.

Beyond the first type error, the most important properties of a test is its power $\beta$ which measures its ability to successfully detect a heterogeneous region. Estimates $\hat{\beta}$ for the univariate tests are the rejection rates for non-homogeneous UR in Table 1. One can see that the univariate HW test has the highest $\hat{\beta}$ among all the presented tests. This is not a surprise since the regions are generated from a lognormal distribution (see section 4.1). Indeed, the lognormal distribution has many similarities with the generalized extreme value distribution and the generalized Pareto distribution, both particular cases of the four parameter Kappa distribution. However, note that the M test has power estimates $\hat{\beta}$ almost as high as the estimates of the HW test. While the B test has lower power estimates $\hat{\beta}$, the power estimates of the Bc test are as good as those of the HW test. Indeed, the difference between the power estimates of the HW and Bc tests are never larger than 1.2%. The latter have even a better power when the region is bimodal. Thus, even considering that the choice of the four parameter Kappa distribution is justified in this case, the M and Bc tests perform as well as the HW test.

It was pointed out in the introduction that the drawbacks of the HW test would be more important in the multivariate framework. This is actually confirmed by the power estimates of multivariate tests in Table 2. Indeed, in this context, the HW test has much lower $\hat{\beta}$ than the M, B and Bc tests. This shows that the estimation of several parameters creates a large variability, visible in the simulated $V^{(b)}$ values. Once again, the M test shows almost the best performance since it has a larger power for several types of multivariate regions. The Bc test is the second most powerful test since it often has the highest power estimate $\hat{\beta}$. Finally, the B test has lower power estimates than the Bc test, confirming the superiority of the latter.

The Ys and Yr tests show very low $\hat{\beta}$. This poor power is caused by the additional level of variability introduced by the Pólya procedure. In the Ys test, the Pólya resampling is carried out independently for each site which creates variability between the simulated sites $\mathbf{x}^{(b)}$. In the case of Yr, generating a Pólya sample for each region introduces an additional level of resampling. The variability created by the additional level is added to the simulated $V^{*(b)}$. Thus, the Pólya resampling is not relevant in the hypothesis testing context.

Tables 1 and 2 also provide insights about the influence of the region size, *i.e.* number of sites $N$ in the region. First, note that $N$ does not seem to have an influence when the region is actually homogeneous. Second, Chebana and Ouarda (2007) noted that the power of the HW test increases with the number of sites in the region, which is confirmed by Tables 1 and 2. This is also true with the nonparametric tests where an increase in the power estimates is visible when $N$ becomes larger according to Tables 1 and 2. To explain this, note that when $N$ increases, the length of the pooled sample $\mathbf{x}^{tot}$ also increases and its distribution becomes more precise. When drawing the sites $x^{(b)}$ (or $x^{*(b)}$) from $\mathbf{x}^{tot}$ (for all methods), there is less variability in their distribution which usually results in lower simulated $V^{(b)}$ (or $V^{*(b)}$) values. This is illustrated in the univariate framework by the distribution of the $P=500$ computed rejection thresholds (the 95$^{th}$ percentile of the $V^{(b)}$ or $V^{*(b)}$ distribution) shown in Figure 2 with $N$ = 15, 20 and 30. It is clear that the median and standard deviation of the simulated $V^{(b)}$ values decrease when $N$ increases. This is less clear in Figure 3 which is equivalent to Figure 2 but for the multivariate framework. The reduction of median and standard deviation of the threshold is still visible for the HW, B and Bc tests when $N$ increases. However, the opposite is happening for the M, Ys and Yr tests. For the Ys and Yr tests, it seems that the extra variability caused by the extra level of resampling is worse when $N$ increases. Hence, increasing the number of sites indeed improve the tests power, except for the Ys and Yr tests which create variability from one site to another.

Table 3 shows the mean computation time for performing each resampling method once (HW, M, B and Y, regardless of the little variation inside each framework). It is shown that there is a huge gain in computation time with the nonparametric framework. This is especially due to the fitting of the Kappa distribution which uses an iterative optimisation algorithm. This kind of algorithm is often time consuming and does not always converge. For instance, during the simulation study, no result could be obtained for the HW test for around 5% of cases due to the non-convergence of the algorithm. Thus when it happens to real data, this could be a real problem to the application of the test.

## 5. Application to the Côte-Nord case

The present section deals with the application of the proposed nonparametric homogeneity test to a real world dataset. The dataset represents the volume and peak of flood events in the Côte-Nord region of the province of Quebec, Canada. The data is drawn from $N = 26$ sites with record lengths ranging from 14 to 48 years. The geographic location of the stations is shown in Figure 4. More details about this dataset can be found in Chebana et al. (2009). We consider the analysis in the univariate case for the variables volume and peak separately as well as jointly in the bivariate case.

In the first step of the RFA we discard the discordant sites from the data set by using the discordancy test of Hosking and Wallis (1993) for the univariate case and the test of Chebana and Ouarda (2007) for the multivariate one. After the application of the discordancy test in the Côte-Nord region (see, Chebana et al. 2009), Site 2 was removed from all the analyses because of its very high L-CV statistics for both the volume and peak. In addition, Site 16 was removed from the region for the univariate analysis of the volume because of its extremely low L-CV and L-Skewness statistics. Finally, Site 21 was discarded for the bivariate analysis, because it combines low L-CV for both volume and peak characteristics and a high skewness for the volume.

Table 4 shows the homogeneity test results for the volume and peak and the bivariate analysis of the region from which the discordant sites have been discarded. At the significance level $\alpha = 5\%$, the original HW test and the nonparametric tests agree to accept the null hypothesis of homogeneity in the volume and in the bivariate case. However, for the peak, the original HW test largely rejects the null hypothesis in opposition to all nonparametric tests which accept it.

Figure 5 allows to make an interpretation of the agreement and disagreement of the HW test and the nonparametric tests for the volume and the peak respectively. It shows the Q-Q plots of the volume and peak distributions versus the Kappa distribution for two sites. The two sites shown (Site 1 and Site 11) represent well the fit of the other sites. It can be seen that the fit of the Kappa distribution is not perfect for both the volume (Figure 5a and 5b) and the peak (Figure 5c and 5d). Note that the fit is especially bad at the extremes for the peak, and that it is for the peak that the nonparametric tests disagree with the HW test. It shows that the prior assumption of a parametric distribution can be wrong, especially for the margins in the RFA case since Chebana et al. (2009) showed that the fit was fairly good for the copula. Therefore, in the present case, the nonparametric tests seem more reliable to make inference concerning the homogeneity of the region.

## 6. Conclusion

Given the drawbacks of the HW test, this paper proposes to introduce nonparametric procedures to homogeneity testing in the RFA context. The improvements are related to the simulation of homogeneous regions and to the definition of the decision threshold. To overcome these limitations, several nonparametric methods are compared: permutation methods, bootstrap, and Pólya resampling. Moreover, a p-value is computed to decide whether the hypothesis of homogeneity should be rejected.

A simulation study is carried out in order to evaluate and compare the performances of the tests. The performances were evaluated according to two fundamental properties: first type error and power. The former should be close to the chosen significance level $\alpha$ for the test to be exact. As expected, the M test was found exact as well as the B test in the univariate framework and the Bc test in the multivariate one. The univariate HW test was found to have a very higher first type error while its multivariate version was found to be conservative. This indicates that the critical region of this test was not accurate enough. We also found that the Pólya based tests (Yr and Ys) were too conservative.

As important as the accuracy of the first type error is the power of the test. In the univariate framework, we found that the HW test has a power as high as the M and Bc tests. Note however that since the HW test's first type error was too high, its power is probably overestimated. In the multivariate framework, the M, B and Bc tests have much higher power than the HW test. This confirms that the drawbacks of the HW test are an actual limitation of the test when the number of variables grows. Among the three cited tests, the most powerful are the M and Bc tests which have very close powers. The Ys and Yr tests show the lowest power among all the tests. This is explained by the extra level of resampling in their procedures. Thus, to the Pólya resampling is not recommended in the present hypothesis testing context. The real life application shows that there are cases where the HW test can reject the null hypothesis of homogeneity while its nonparametric counterparts accept it easily. However it also shows that the Kappa distribution assumption of the margins is not well respected, which then justifies the nonparametric framework.

To conclude, the L-moment homogeneity test requires a nonparametric framework to increase its applicability and performances. This is provided by the bootstrap and permutation methods which allow increasing the power of the test in the multivariate case where modelling limits are noticeable. Among all the procedures compared in this paper, we would recommend the M test based on permutations. It is indeed an exact test and as such, the most powerful among the proposed tests. Moreover, permutation-based methods are especially well suited for testing a hypothesis such as homogeneity. The M test thus results in an increase of the power of the homogeneity test and its applicability since no user intervention is needed and no numerical optimisation is performed. It should, in the end, increase the performances of the RFA.

Table 1: Simulation results on region UR. This table presents the obtained rejection rates at significance level $\alpha = 5\%$ for each test and several kinds of generated regions. Values of the heterogeneity and rejection rates are expressed in percentages. When the region is heterogeneous, the bold value indicates the higher rejection rate.

| Type | $\gamma_\tau$ (%) | $\gamma_{\tau_3}$ (%) | N | HW (%) | M (%) | B (%) | Bc (%) | Ys (%) | Yr (%) |
|---|---|---|---|---|---|---|---|---|---|
| Hom. | - | - | 15 | 6.2 | 4.0 | 5.0 | 6.4 | 3.4 | 3.8 |
| | | | 20 | 6.1 | 5.6 | 4.8 | 6.6 | 3.4 | 4.0 |
| | | | 30 | 6.8 | 5.0 | 5.4 | 6.8 | 4.2 | 4.2 |
| Lin. | 37.5 | 200 | 15 | **49.5** | 49.2 | 45.8 | 48.2 | 38.2 | 41.2 |
| | | | 20 | 57.7 | 54.6 | 55.2 | **59.2** | 50.0 | 50.0 |
| | | | 30 | **72.0** | 69.8 | 66.8 | 71.8 | 62.4 | 63.0 |
| Bim. | 37.5 | 200 | 15 | 95.6 | 94.4 | 95.0 | **95.8** | 92.0 | 91.0 |
| | | | 20 | **98.4** | 98.0 | 98.2 | 98.0 | 96.8 | 96.8 |
| | | | 30 | 99.8 | 99.4 | **100** | **100** | 99.8 | 99.4 |

Table 2: Simulation results on region MR. This table presents the obtained rejection rates at significance level $\alpha = 5\%$ for each test and several kinds of generated regions. Values of the heterogeneity and rejection rates are expressed in percentages. When the region is heterogeneous, the bold value indicates the higher rejection rate.

| Type | $\gamma$ (%) | N | HW (%) | M (%) | B (%) | Bc (%) | Ys (%) | Yr (%) |
|---|---|---|---|---|---|---|---|---|
| Hom. | - | 15 | 3.2 | 3.0 | 3.0 | 4.0 | 1.0 | 1.0 |
| | | 20 | 2.0 | 5.0 | 3.0 | 4.0 | 2.0 | 2.0 |
| | | 30 | 1.0 | 4.0 | 3.0 | 7.0 | 3.0 | 1.0 |
| Lin. Mar. | 30 | 20 | 6.2 | 16.0 | 9.0 | **18.0** | 4.0 | 4.0 |
| | 50 | | 33.0 | 44.0 | 36.0 | **48.0** | 15.0 | 13.0 |
| Lin. Dep. | 30 | 20 | 4.2 | 9.0 | 5.0 | **10.0** | 2.0 | 3.0 |
| | 50 | | 7.3 | **28.0** | 13.0 | 27.0 | 4.0 | 3.0 |
| Lin. Co. | 30 | 20 | 16.3 | 28.0 | 18.0 | **32.0** | 11.0 | 6.0 |
| | 50 | | 67.4 | **82.0** | 75.0 | **82.0** | 72.0 | 65.0 |
| | | 30 | 79.0 | **95.0** | 91.0 | 93.0 | 82.0 | 71.0 |
| Bim. Mar. | 30 | 20 | 21.9 | 41.0 | 27.0 | **45.0** | 22.0 | 17.0 |
| | 50 | | 81.4 | 90.0 | 86.0 | **93.0** | 83.0 | 78.0 |
| Bim. Dep. | 30 | 20 | 7.4 | 18.0 | 12.0 | **20.0** | 7.0 | 5.0 |
| | 50 | | 30.3 | **74.0** | 50.0 | 64.0 | 47.0 | 39.0 |
| Bim. Co. | 30 | 20 | 75.0 | **85.0** | 77.0 | 83.0 | 67.0 | 61.0 |
| | 50 | | **100.0** | **100.0** | **100.0** | **100.0** | **100.0** | 99.0 |

Table 3: Mean time of multivariate test execution in minutes.

| Resampling | HW | M | B | Y |
|---|---|---|---|---|
| Mean time | 12'38'' | 0'11'' | 0'35'' | 0'41'' |

Table 4: Homogeneity results of the application to the Côte-Nord data.

| | Discarded sites | Test | $V^{obs}$ | $\mu_{sim}$ | $\sigma_{sim}$ | $H$ | p-value | Decision |
|---|---|---|---|---|---|---|---|---|
| Volume | 2, 16 | HW | 0.0231 | 0.0219 | 0.0032 | 0.5282 | 0.340 | Hom. |
| | | M | | 0.0473 | 0.0074 | - | 1.000 | Hom. |
| | | B | | 0.0470 | 0.0071 | - | 1.000 | Hom. |
| | | Bc | | 0.0451 | 0.0070 | - | 1.000 | Hom. |
| | | Ys | | 0.0481 | 0.0071 | - | 1.000 | Hom. |
| | | Yr | | 0.0474 | 0.0068 | - | 1.000 | Hom. |
| Peak | 2 | HW | 0.0321 | 0.0230 | 0.0035 | 2.6233 | 0.008 | Heter. |
| | | M | | 0.0446 | 0.0062 | - | 0.988 | Hom. |
| | | B | | 0.0454 | 0.0065 | - | 0.982 | Hom. |
| | | Bc | | 0.0402 | 0.0064 | - | 0.890 | Hom. |
| | | Ys | | 0.0452 | 0.0070 | - | 0.986 | Hom. |
| | | Yr | | 0.0448 | 0.0067 | - | 0.970 | Hom. |
| Bivariate | 2, 21 | HW | 0.0296 | 0.0251 | 0.0028 | 1.6071 | 0.066 | Hom. |
| | | M | | 0.0468 | 0.0071 | - | 1.000 | Hom. |
| | | B | | 0.0475 | 0.0066 | - | 1.000 | Hom. |
| | | Bc | | 0.0451 | 0.0060 | - | 1.000 | Hom. |
| | | Ys | | 0.0479 | 0.0069 | - | 0.998 | Hom. |
| | | Yr | | 0.0472 | 0.0064 | - | 1.000 | Hom. |

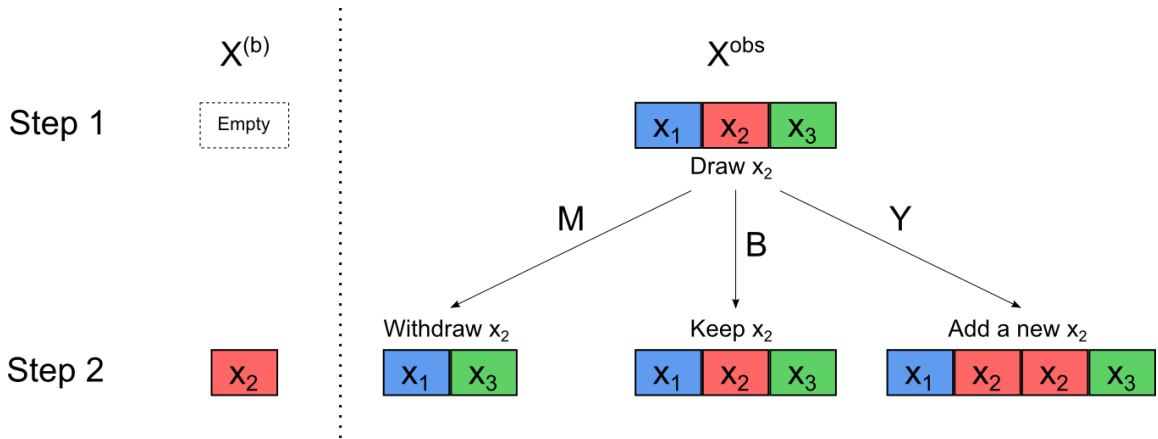

**Figure 1: Illustration of the three resampling techniques used in the present work. To the left of the dotted line is the evolution of the simulated sample and to the right is the evolution of the observed one. M: permutation, B: traditional bootstrap, Y: Pólya resampling.**

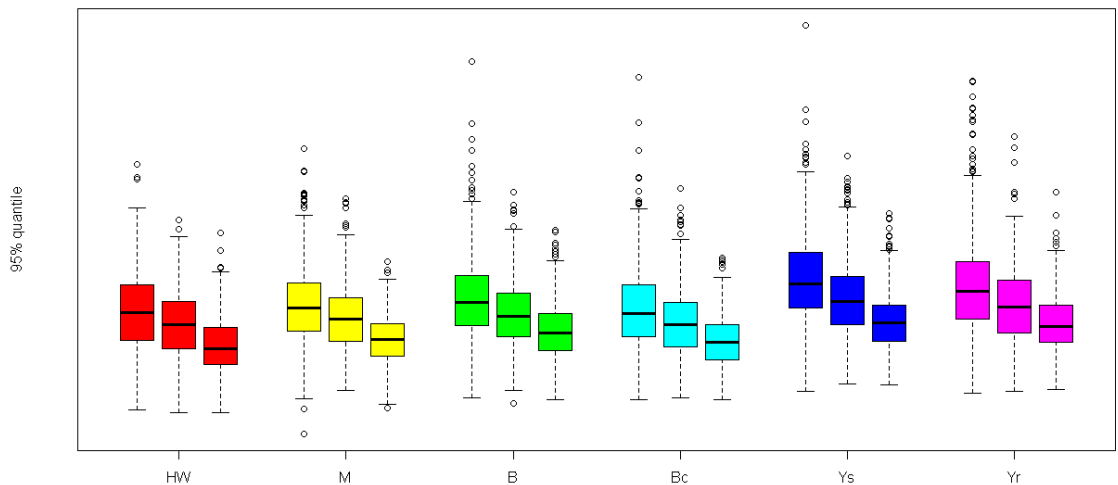

**Figure 2: Boxplots showing the repartition of the $P=500$ 95$^{th}$ centile of $V^{(b)}$ or $V^{*(b)}$ distribution obtained with simulations of UR heterogeneous regions with $\gamma_\tau = 37.5\%$ heterogeneity on the L-CV and $\gamma_{\tau_3} = 200\%$ heterogeneity on the L-skewness. Each color corresponds to a test and for each test, the first box corresponds to a region of size $N=15$, the second corresponds to $N=20$ and the third corresponds to $N=30$.**

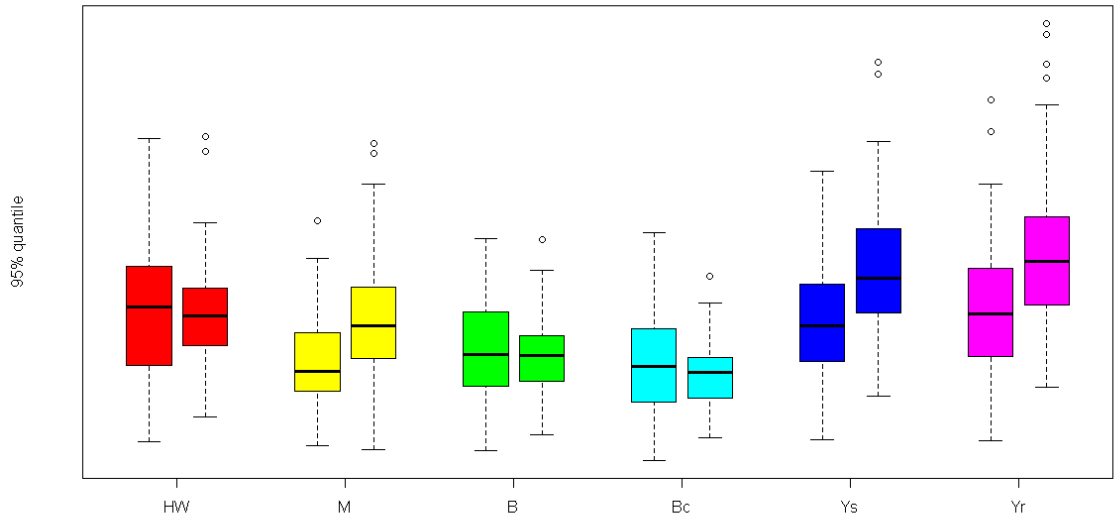

Figure 3: Boxplots showing the repartition of the $P=100$ 95$^{\text{th}}$ centile of $V^{(b)}$ or $V^{*(b)}$ distribution obtained with simulations of MR completely linearly heterogeneous regions with $\gamma = 50\%$ heterogeneity on the margin and the dependence parameter. Each color corresponds to a test and for each test, the first box corresponds to a region of size $N = 20$ and the second corresponds to $N = 30$.

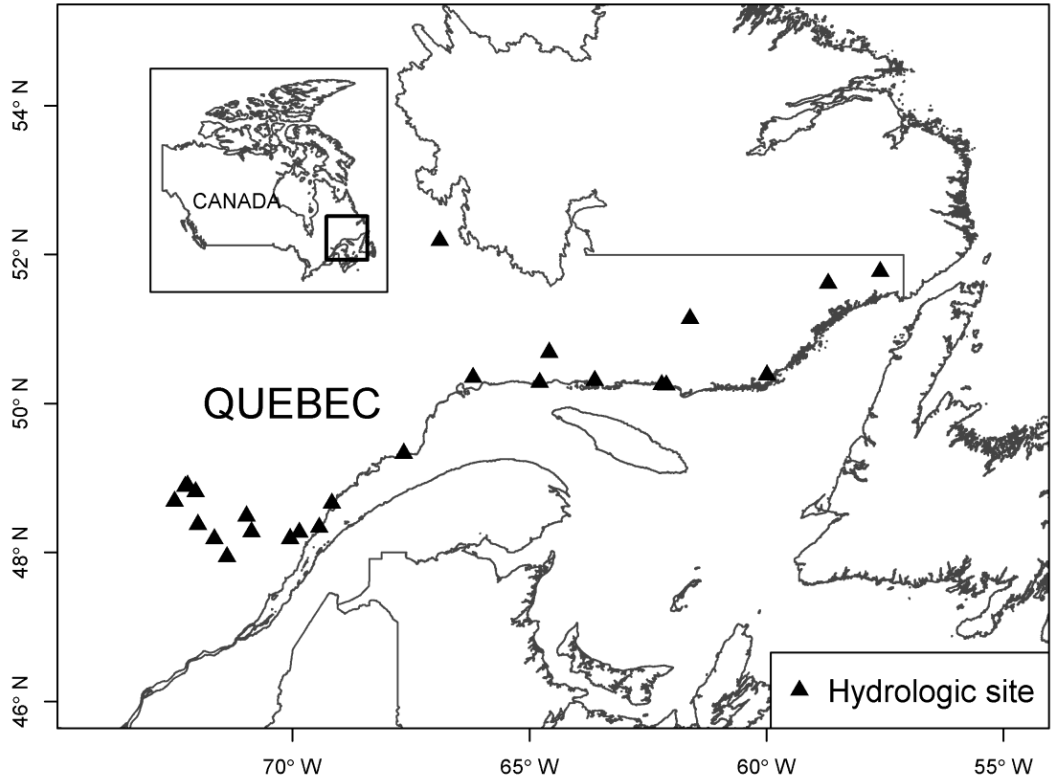

Figure 4: Geographical location of the $N = 26$ considered sites in the Côte-Nord case study.

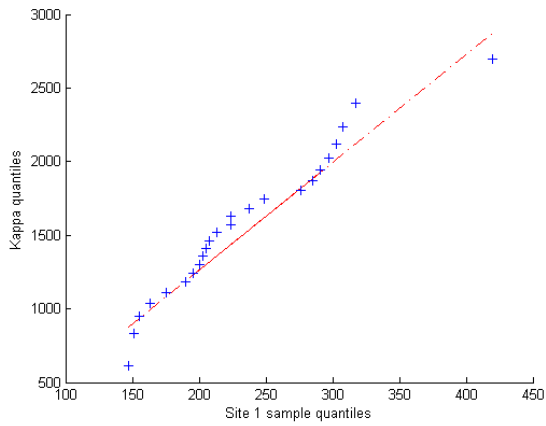
a) Site 1: volume

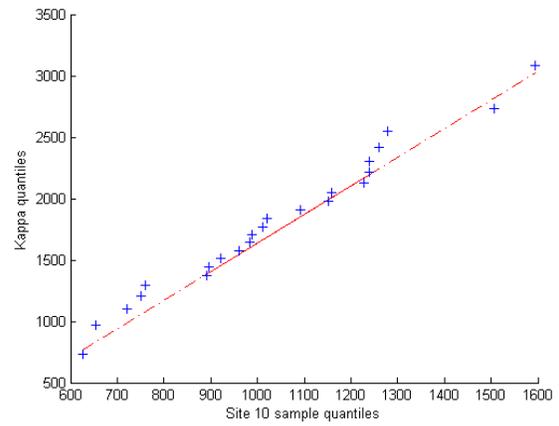
b) Site 11: volume

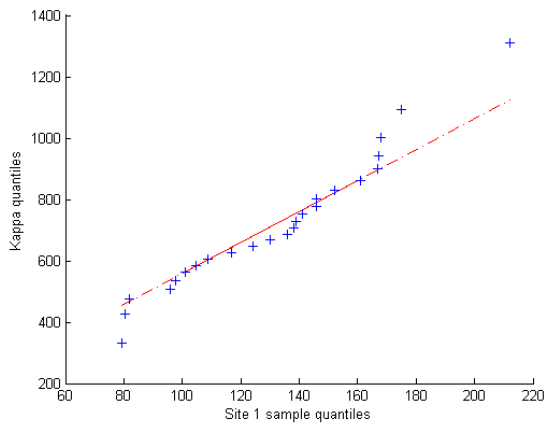
c) Site 1: peak

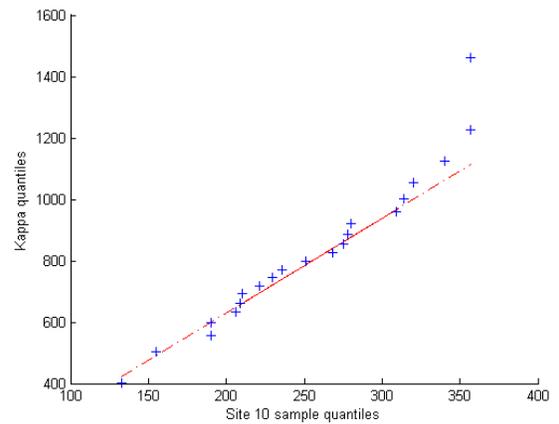
d) Site 11: peak

Figure 5: Examples of Q-Q plots showing sample univariate distribution versus the Kappa quantiles for sites 1 and 11. The parameters of the Kappa distribution from which the quantiles are drawn are the ones estimated from the data.